\documentclass{article}
\usepackage[T1]{fontenc}
\usepackage[utf8]{inputenc}
\usepackage{ismir} % Assuming this is a custom package
\usepackage{amsmath}
\usepackage{cite}
\usepackage{url}
\usepackage{graphicx}
\usepackage{amssymb}
\usepackage{lipsum} % For dummy text
\usepackage{multicol} % For multi-column layout
\usepackage{float} % For H option in table*
\usepackage{multirow}
\usepackage{algorithm}
\usepackage{algpseudocode}
\usepackage{array}
\newcolumntype{P}[1]{>{\centering\arraybackslash}p{#1}}
\newcolumntype{M}[1]{>{\centering\arraybackslash}m{#1}}
\usepackage{ragged2e}
\usepackage{color}
\usepackage{caption}
\usepackage{tabularray}
\usepackage{booktabs}
\usepackage{amssymb}% http://ctan.org/pkg/amssymb
\usepackage{pifont}% http://ctan.org/pkg/pifont
\newcommand{\cmark}{\ding{51}}%
\newcommand{\xmark}{\ding{55}}%
\usepackage{enumitem}
\usepackage{hhline}

\title{Wavespace: A Highly Explorable Wavetable Generator}

\multauthor
{Hazounne Lee$^1$ \hspace{1cm} Kihong Kim$^2$ \hspace{1cm} Sungho Lee$^1$ \hspace{1cm} Kyogu Lee$^1$}
{
\\ $^1$ Department of Intelligence and Information, Seoul National University\\
$^2$ Department of Mathematics, Kyungpook National University\\
{\tt\small \{hazounne, sh-lee, kglee\}@snu.ac.kr, kimgihong2510@knu.ac.kr}
}

\def\authorname{Hazounne Lee, Kihong Kim, Sungho Lee, Kyogu Lee}

\usepackage[bookmarks=false,pdfauthor={\authorname},pdfsubject={\papersubject},hidelinks]{hyperref}

\begin{document}
\maketitle
\begin{abstract}
Wavetable synthesis generates quasi-periodic waveforms of musical tones by interpolating a list of waveforms called wavetable.
As generative models that utilize latent representations offer various methods in waveform generation for musical applications, studies in wavetable generation with invertible architecture have also arisen recently.
While they are promising, it is still challenging to generate wavetables with detailed controls in disentangling factors within the latent representation.
In response, we present \texttt{Wavespace}, a novel framework for wavetable generation that empowers users with enhanced parameter controls.
Our model allows users to apply pre-defined conditions to the output wavetables. We employ a variational autoencoder and completely factorize its latent space to different waveform styles.
We also condition the generator with auxiliary timbral and morphological descriptors.
This way, users can create unique wavetables by independently manipulating each latent subspace and descriptor parameters.
Our framework is efficient enough for practical use; we prototyped an oscillator plug-in as a proof of concept for real-time integration of \texttt{Wavespace} within digital audio workspaces (DAWs).
%Our open-source code is available at: \texttt{https://github.com/hazounne/Wavespace}.
\end{abstract}
\section{Introduction}
\label{sec:introduction}
\begin{figure}[t]
 \centerline{
 \includegraphics[width=1\columnwidth]{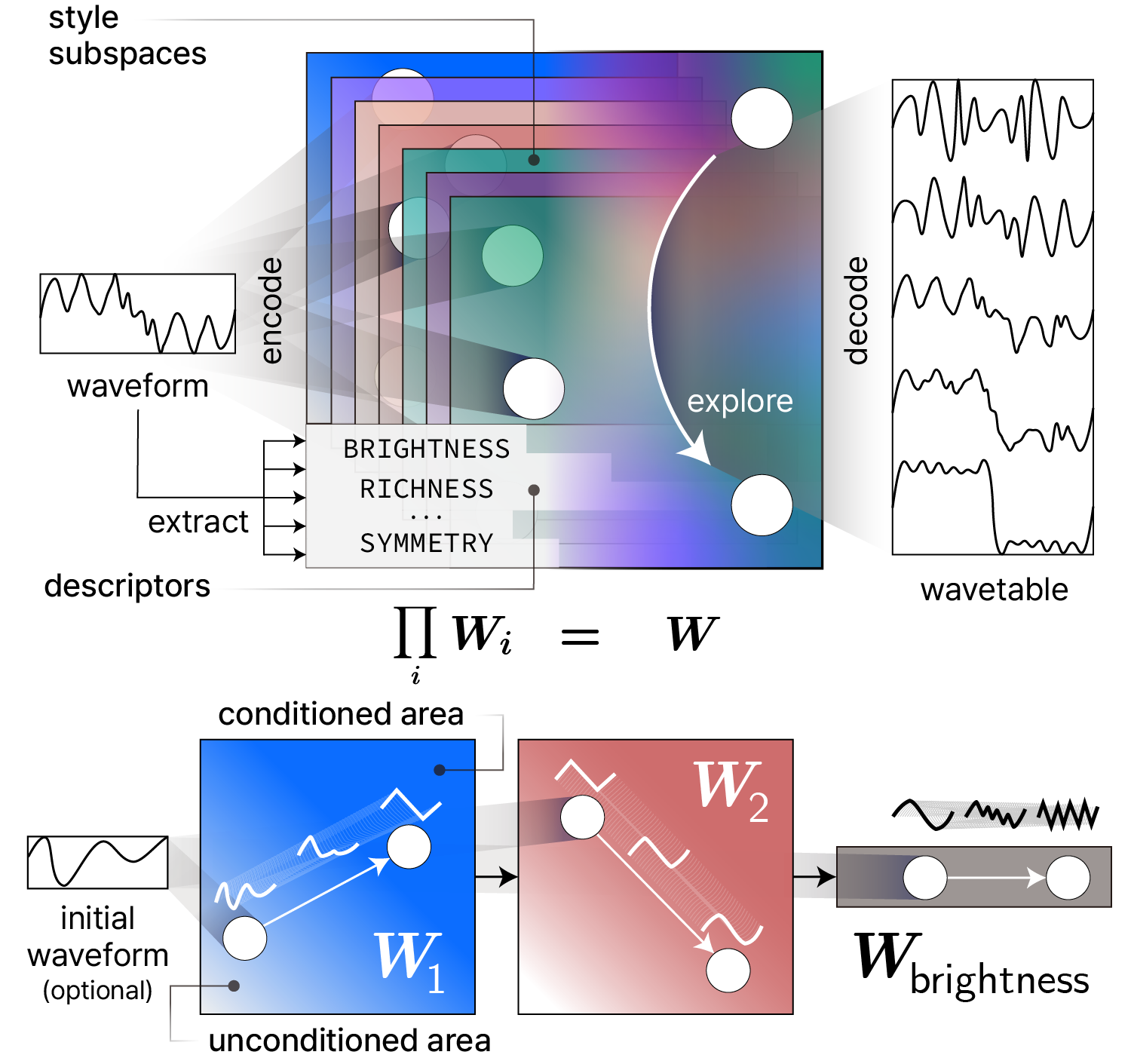}}
 \caption{The \texttt{Wavespace} framework enables waveform creation through subspace encoding and interpolation. Each subspace latent is adjustable independently.}
 \label{fig:fig1}
\end{figure}
%Introducing wavetable synthesis
Among many real-time synthesis methods of musical tones, wavetable synthesis has been widely used to generate quasi-periodic waveforms\cite{bristow1996wavetable:01}.
Wavetable synthesis retrieves single-cycle waveforms from a wavetable. Compared to other sound synthesis techniques, this method empowers users to create more general or evolving waveforms by interpolation of waveforms with efficient memory usage \cite{franck2012higher}. 
While the previous restriction of computation and memory usage is highly mitigated, it is still recognizably used in applications for musical tone synthesis with a wide range of sound textures \cite{horner1997comparison, maher2005wavetable}.

%Invertible architectures are used for audio latent representation
Meanwhile, researchers have studied waveform generation with deep neural networks \cite{oord2016wavenet, chen2020wavegrad}. One of the widely-used methods utilizes audio latent representation in autoencoder structure\cite{morrison2021chunked, yamamoto2020parallel}, especially for musical tone generation\cite{ engel2017neural, tatar2021latent,  esling2018bridging, tahirouglu2021ganspacesynth}.
%Invertible models are used in musical applications
Research on waveform generation for musical applications with these architectures has been emerging in the last few years\cite{engel2020ddsp, hyrkas2021wavaetable, caillon2020timbre, caillon2021rave, tahirouglu2021ganspacesynth}. %real-time이나 wavetable 아닌거만 cite
%Neural Wavetable generator, commonly VAE is used.
One of the applications is the neural wavetable generator, which synthesizes a series of single-cycle waveforms from the latent space to construct a wavetable\cite{hogberg2023latent, tsugumasa2023}. 
Wavetables can be generated by interpolation of latent vectors, enabling distinctive non-linear blending between waveforms \cite{hantrakul2018neural}. 
This commonly utilizes generative models such as variational autoencoders (VAEs).
%Autoencoder-based oscillator
More recently, studies have used this autoencoder structure for an oscillator with real-time waveform generation. By decoding a waveform for every audio buffer request, the latent space replaces the functionality of the wavetable. %from different latent space parameters.
This method is a novel way of generating quasi-periodic waveforms because the latent space exploration allows obtaining various waveforms with smooth transitions between them\cite{krekovic2021concept}.
It has a similar concept to vector synthesis in the aspect of exploring a given space to generate real-time single-cycle waveforms\cite{keatingsurvey}.
This approach is more general than generating wavetables but is also challenging because of the computation speed restriction. In short, current studies focus on either wavetable generation or oscillator implementation. They aim to enable users to generate waveforms utilizing latent representations.
However, it is challenging for users to select specific waveform styles since the latent space is entangled.% without any regularization or structural prior.

To reduce this gap, we present \texttt{Wavespace}, a highly explorable wavetable generator with complete conditional controls.
\autoref{fig:fig1} represents \texttt{Wavespace} and its instructions for wavetable generation. 
Compared to the previous methods, \texttt{Wavespace} explicitly factorizes its latent space so that each subspace modifies the style of the waveform in a more interpretable manner.
This way, users can understand how each latent space parameter works.
Specifically, we disentangle the latent space into style subspaces. 
Waveforms are encoded to each style subspace, and auxiliary descriptors are algorithmically extracted. 
Users can control the style and descriptor parameters to obtain wavetables from scratch. 
First, they initialize the parameters by exploring the latent space.
Encoding a waveform and extracting descriptors can be another way to do it.
They can adjust the parameters to transform the waveform.
For example, in the style subspace $W_1$ the output waveform morphs to the corresponding style by moving the parameter from the unconditioned area to the conditioned area. General waveforms with subtle changes are obtained by passing through two areas as seen in the style subspace $W_2$.
Increasing the parameter in descriptor $W_{\text{brightness}}$ adds more brightness to the waveform. These parameter manipulations accompany a smooth transition in the waveform.
%Training, model simply
Our model is based on conditional variational autoencoder (CVAE)\cite{sohn2015learning} structure.
The latent space is variationally trained with algorithmically extracted descriptors concatenated.
With \texttt{Wavespace} and its plug-in implementation, our major contributions are as follows:
\begin{itemize}[leftmargin=6mm]
\setlength\itemsep{0em}
 \item We introduce \texttt{Wavespace}, a novel framework for flexible and controllable wavetable generation from factorized latent space.
 \item We propose a method encompassing wavetable generation and a real-time autoencoder-based oscillator with complete conditional controls.
 \item We build an audio plug-in based on \texttt{Wavespace} with a user interface and show its real-time use in the CPU.
\end{itemize}

\section{Background}
\subsection{Wavetable Synthesis}
Wavetable synthesis employs fixed-size single-cycle waveforms $x_1, \cdots, x_M \in \mathbb{R}^N$ where $M$ and $N$ denote the number of waveforms and their length in samples, respectively. 
We stack these waveforms to form a so-called ``wavetable" $\textbf{X} \in \mathbb{R}^{M\times N}$.
Then, we can generate a signal $s[n]$ from the wavetable $\mathbf{X}$ with a read operation $\Phi$ given as follows,
\begin{equation}
    s[n]= \Phi(\mathbf{X}, \tilde{i}[n], \tilde{j}[n])
\end{equation}
where $\tilde{i}[n]$ and $\tilde{j}[n]$ are row and column indices, respectively. 
The row index $\tilde{i}[n]$ selects the waveform to use. It is typically a time-varying signal, which changes or ``morphs'' the shape of waveform across time. 
The column indices $\tilde{j}[n]$ are determined by the time-varying (instantaneous) fundamental frequency $f_0[n]$ and given as follows, 
\begin{equation}
    \tilde{j}[n]=\Bigg[\frac{N}{f_s} \sum_{m=0}^n f_0(m) \Bigg] \;\% \;N.
\end{equation}
where $f_s$ and $\%$ denote a sampling rate and modulo operation, respectively. When both indices are integers, we can obtain the desired sample by reading the table element: $s[n] = \mathbf{X}[\tilde{i}[n], \tilde{j}[n]]$.
Otherwise, the read operation $\Phi$ interpolates the nearest samples in the wavetable $\mathbf{X}$, e.g., in a (bi-)linear manner.

\subsection{Neural Wavetable Generators}
Instead of relying on a fixed and pre-defined wavetable $\mathbf{X}$, 
existing works employ an autoencoder framework, using a pair of encoder $\mathcal{E}$ and decoder $\mathcal{D}$ to obtain a generative model of single-cycle waveforms. 
Once trained, one can obtain a novel wavetable $\mathbf{X}$ by decoding a series of latent vectors $\mathbf{w}_1, \cdots, \mathbf{w}_M \in W$ as follows,
\begin{equation}
    \mathbf{X} = [{x}_1, \cdots, {x}_M] = [\mathcal{D}({\mathbf{w}}_1), \cdots, \mathcal{D}({\mathbf{w}}_M)].
\end{equation}
\cite{hantrakul2018neural} shows the basic concept of the neural wavetable generator. The authors demonstrate blending two single-cycle waveforms by latent interpolation, allowing for the creation of intermediate waveforms. However, the method's reliance on a large model hinders real-time generation.

There are studies for wavetable generators focusing on real-time autoencoder-based oscillators.
\cite{krekovic2021concept} introduce this oscillator concept with VAE architecture. While its method allows for generating various waveforms with smooth transitions, the entangled latent space makes it difficult and time-consuming for users to navigate and find the specific sounds they desire.
\cite{tsugumasa2023} present a method for generating single-cycle waveforms using timbral descriptors. However, this approach requires an input waveform to be morphed, limiting its ability to create waveforms from scratch. The latent space entanglement is the direct cause of this limitation.
We factorize the latent space and integrate the separate wavetable generation methods into \texttt{Wavespace} framework with sufficient parameters to craft waveforms.
%model introduction: factorized latent space -> conditional control

\section{Method}
We build upon the architecture of our baseline described in \cite{tsugumasa2023}. The baseline model has a CVAE structure with both time-domain input and output, employing three timbral descriptors. It consists of 1D convolutional and transposed convolutional layers.
%This follows the traditional additive synthesis method.

\subsection{Learning Latent Subspaces}
\begin{figure}[t]
 \centerline{
 \includegraphics[width=1\columnwidth]{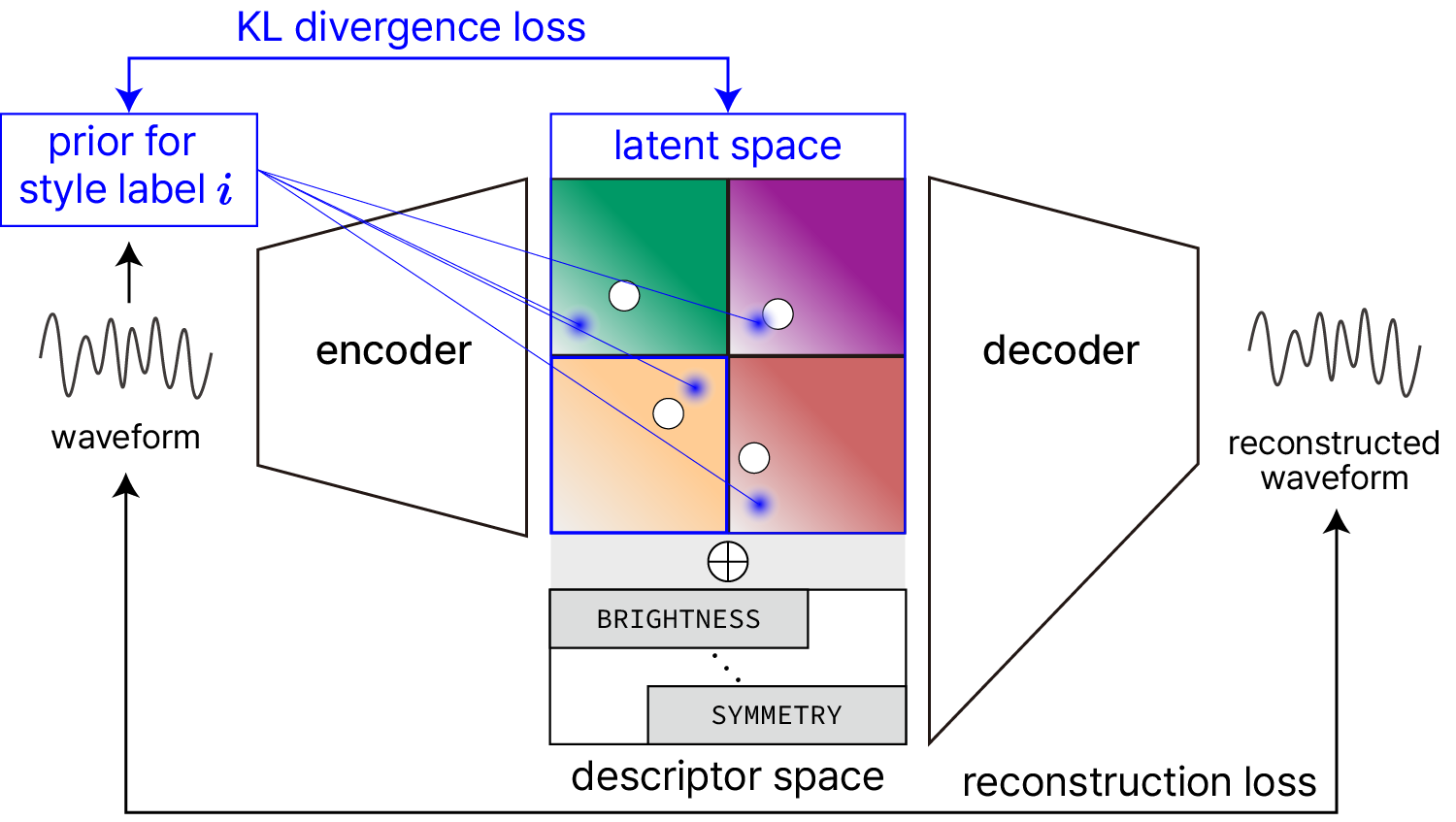}}
 \caption{Our model's training method learns latent subspaces. Priors depend on the target style label. The prior within the blue-bordered subspace has a unique position.}
 \label{fig:fig2}
\end{figure}

The parameter space $W$ is a product of the style latent space $W_S$ and the descriptor space $W_D$. $S$ and $D$ refer to a set of styles and descriptors.
% A combination of parameters is decoded as a relevant waveform.
We can break down the parameter space as follows:
\begin{equation} \label{eu_eq4}
W = W_S \times W_D = \prod_{s \in S}W_s \times \prod_{d \in D}W_d
\end{equation}
% The target style label $i$, and descriptor parameter $\mathbf{d}$ are the conditions to be considered.
We learn latent subspaces $W_s (s \in S)$ to factorize the latent space, inspired by \cite{klys2018learning}. The study replaces condition concatenation with another variationally trainable latent space. $\mathbf{s}$ and $\mathbf{d}$ are style and descriptor parameters.
For $\mathbf{s}_j \in W_j$, the prior in $j$th latent subspace,  $p_i(\mathbf{s}_j)$, differs from two situations; the label corresponds to the subspace ($i=j$) or otherwise ($i\neq j$). The subspace prior switches as
 \begin{equation} \label{eu_eq2}
p_i(\mathbf{s}_j) \sim \begin{cases} 

\mathcal{N}(\mu_{0j},\sigma_{0j}^2), \quad i \neq j\\ 

\mathcal{N}(\mu_{1j},\sigma_{1j}^2), \quad i = j

\end{cases},
\end{equation}
for $i,j = 0, 1, 2, .., N-1$. $\mu_{0j}$, $\mu_{1j}$, $\sigma^2_{0j}$, and $\sigma^2_{1j}$, which are the mean and variance of Gaussian distribution. Note that only the $i$th subspace has a different subspace prior, i.e. $p_i(\mathbf{s}_i)$. We had experiments $\mu_{0j}$ and $\mu_{1j}$  both fixed and learned from random initialization. A jointed prior for target $i$ is obtained as
 \begin{equation} \label{eu_eq3}
p_i(\mathbf{s}) = \prod^{N-1}_{j=0} p_i(\mathbf{s}_j)\sim \mathcal{N}(\mu_i, \sigma^2_i).
\end{equation}
The assumption on this statement is that $p_i(\mathbf{s}_j)$ is mutually independent for every $j$.
% Each distribution $q_\phi(\mathbf{s}_i|\mathbf{x})$ subspace learns from jointed priors $p_i(\mathbf{s})$.
This method disentangles the style subspaces and integrates them into a shared latent space. Then, we optimize the variational evidence lower bound as an objective given as follows,
\begin{equation}
\begin{split}
\mathbb{E}_{q_{\phi}(\mathbf{s}|\mathbf{x}, \mathbf{d})} \left[ \log p_{\theta}(\mathbf{x}|\mathbf{s}, \mathbf{d}) \right]
- D_{\text{{KL}}}\left( q_{\phi}(\mathbf{s}|\mathbf{x}, \mathbf{d}) || p_i(\mathbf{s}) \right).
\end{split}
\label{eq14}
\end{equation}
This loss aims to generate waveforms according to the style latent space parameters, by learning properly both $q_\phi(\mathbf{s}|\mathbf{x},\mathbf{d})$ and $p_\theta(\mathbf{x}|\mathbf{s},\mathbf{d})$. The overall scheme of learning latent subspaces is illustrated in \autoref{fig:fig2}. 

\subsection{Descriptor Extraction}
\label{subsubsec:3.2.4}
% \cite{krekovic2016algorithm} introduced sound synthesizers with adjectives by mapping them to audio features extracted with algorithms.
 We obtain timbral descriptors from relevant spectral audio features based on the method utilized in \cite{krekovic2022deep, tsugumasa2023}. Readers can refer to \cite{krekovic2016algorithm} and references therein for the relation between audio features and timbral descriptors. Moreover, we add morphological descriptors which are obtained from waveform features.
Although these features may not be independent of the timbre, they allow particular morphs in sound texture in distinctive ways.
For some descriptors, we extract features and then compress their range with the following function $\sigma$ defined as follows,
\begin{equation}
    \sigma(d) = \frac{\log(d\times (e^k-1)+1)}{k}
\end{equation}
where the hyperparameter $k$ is empirically set to $5.5$. Below are the details of the descriptors.
\begin{itemize}[leftmargin=6mm]
\setlength\itemsep{0em}
    \item \textbf{Brightness ---} The spectral centroid refers to the center of mass of the spectrum indicating the balance between low and high-frequency energy. We relate the brightness of the sound to the spectral centroid as
        \begin{equation}\label{eq:brightness}
            d_\mathrm{B}(x) = \sigma \Bigg(\sum^{N/2}_{k=0} k|X[k]|^2\Bigg)
        \end{equation}
    where $X$ denotes the discrete Fourier transform (DFT) of the single-cycle waveform $x$.
    \item \textbf{Richness ---} Similar to brightness, we relate harmonic spread to the richness of the sound. % and applied the same scaling method.
        \begin{equation}
            d_\mathrm{R}(x) = \sigma \Bigg(\sum^{N/2}_{k=0} |(k-C)X[k]|^2\Bigg)
        \end{equation}
    where $C$ is the spectral centroid defined as above but before the compression $\sigma$.
    \item \textbf{Fullness ---} This feature is opposite to the warmness, which is discovered from sounds without even overtones. Thus, we use the power ratio between the odd overtones and the total overtones.
        \begin{equation}
            d_\mathrm{F}(x) = 1-\frac{\sum_{k=0}^{N/4-1} |X[2k+1]|^2}{\sum_{k=0}^{N/2} |X[k]|^2}. 
        \end{equation}
    \item \textbf{Undulation ---} 
    We assume that the difference of a signal is employable for distinctive timbral manipulation.
    We use the scaled absolute of difference to quantify the ``zigzagness'' of a waveform.
        \begin{equation}
            d_\mathrm{U}(x) = \frac{1}{N-1} \cdot \sigma \Bigg( \sum_{n=0}^{N-2} |x[n+1]-x[n]|\Bigg).
        \end{equation}
    \item \textbf{Symmetry ---} We employ the angle of the sum of all the magnitude components to improve the model to distinguish between two waveforms that have the same amplitude but different phases.
        \begin{equation}
            d_\mathrm{S}(x) = \angle \Bigg( \sum_{k=0}^{N/2} X[k] \Bigg).
        \end{equation}
\end{itemize}
Our hypothesis is that concatenating these descriptors would improve the quality of reconstruction and latent representation. This is because we provide additional information about the waveform in various ways, which may not be captured from the low-dimensional subspaces. 

\subsection{Architecture Details}
The encoder consists of six 1D convolution blocks, each including Leaky ReLU of negative slope of 0.2 and batch normalization, followed by the final linear layer and reparametrization, outputting values for style latent space. Descriptors are concatenated to the latent space and then passed to the decoder. The decoder comprises 6 blocks of upsampling and residual layers, in which a Transposed 1D convolution and three 1D convolution layers are used, respectively. The decoder outputs raw waveform. The final waveform is normalized to the power of 1 with any constant offset eliminated.

%We enumerate descriptors in this order if not specified.

% \subsection{Additive Synthesis}
%  Additive synthesis is the first technique widely used for sound synthesis of audio in computer music by adding sine waves together.\cite{maher1991sinewave}

%  with many complex synthesizers adopting this approach to construct one or more oscillators within the device.[CITE] Additive synthesis has gained prominence due to its ability to represent complex harmonic changes, making it a preferred choice for creating rich and dynamic sounds.
%  the waveform can be obtained from amplitude $A_i$ and phase $\phi_i$ of the $i$th overtone in $i=1,...,N-1,N$ as \begin{equation} \label{eq1}
% s(t) = \sum_{i=1}^{N} A_i \sin\left(2\pi f_it \cdot T_s + \phi_i\right),
% \end{equation}where $T_s=\frac{1}{f_s}$ denotes sample period and $\phi_i$ is the phase of $n$-th overtone. In the additive synthesizer scheme, the DC offset is commonly disregarded and set to 0. To ensure adaptability to the additive synthesizer pipeline, we structured the decoder block to directly output magnitudes separated as amplitude and phase. This allows for waveform synthesis via inverse discrete Fourier transform (IDFT) to obtain the final waveform.\\
\section{Experiments}
\subsection{Datasets}
According to \cite{krekovic2021concept}, the model capacity relies heavily on the dataset used for training. To evaluate the model performance with different datasets we utilize two different datasets. We train our baseline architecture with our dataset and compare it to our model. 
\begin{itemize}[leftmargin=6mm]
\setlength\itemsep{0em}
    \item \textbf{Serum ---} We conduct experiments using wavetables extracted from the Serum VST plugin, a widely recognized standard in the field of wavetable synthesizers\footnote{\url{https://xferrecords.com/products/serum/}}. We select 18 wavetables and assign them to each style. each wavetable has 256 waveforms of 2048 samples.
    \item \textbf{WaveEdit ---}
    we use open-source wavetables provided on WaveEdit Online\footnote{\url{https://waveeditonline.com/}}. We divide the dataset into two styles randomly. Each wavetable has 2572 waveforms of 256 samples.
\end{itemize}
We preprocess the waveforms as follows. First, we resample them to model input size, which is 1024 and 600 for our models and the baseline respectively. We also eliminate any constant offset. Then, the total energy of the waveform is normalized to 1. 
%Due to the tentative reliance of the model performance on the dataset as well as the small volume of data, 
To reduce the variance of the experimental results, we employ $5$-fold cross-validation and report the average results.
%We employ cross-validation by splitting the training and test set by 8 to 2. We then average the test results from 5 experiments in the same settings.

\subsection{Model}
% We test four models which are different in prior learning and model size. 
We use two versions of our model: WS (\textasciitilde1.7M) and WS-S (\textasciitilde111K parameters),  where ``S'' denotes ``small''.
Priors are set to $\mu_{0 i}=(0,0), \mu_{1 i}=(5,5)$, and $\sigma_{0 i}=\sigma_{1 i}=1$ for every $i$. 
We conduct additional experiments of learning $\mu_{i}$ in priors.
In this case, each $\mu_{i}$ is set in hyperspherical coordinates and learns angles to keep $|\mu_{i}|=5$.  We use 2-dimensional style subspaces.

\subsection{Optimization}
%The loss function incorporates conditional loss and waveform $L_1$ norm. Given the $i$th style target,
We draw upon \eqref{eq14} to construct the total loss.
The total loss is formulated as
\begin{equation}
\mathcal{L} = \beta_{1}\mathcal{L}_{\text{s}} + \beta_{2}D_{\text{{KL}}}\left( q_{\phi}(\mathbf{w}|\mathbf{x}, \mathbf{y}) || p_i(\mathbf{w}) \right) 
+ \beta_{3}\mathcal{L}_{\text{w}}.
\end{equation}
Here, \(\mathcal{L}_{\text{s}}\) and \(\mathcal{L}_{\text{w}}\) are the spectral loss and waveform loss respectively.
$\mathcal{L}_w$ is an $L_1$ norm loss for learning the waveform shape, and
$\mathcal{L}_{\text{s}}$ is an $L_2$ norm loss of spectral amplitude.
The first two terms are related to \eqref{eq14} for learning the spectral representation and the encoder distribution.
Because we use Gaussian priors, the reconstruction term $-\mathbb{E}_{q_{\phi}(\mathbf{s}|\mathbf{x}, \mathbf{d})} \left[ \log p_{\theta}(\mathbf{x}|\mathbf{s}, \mathbf{d}) \right]$ requires to be an $L_2$ norm loss. Therefore, we first assign $L_s$ to this term and introduce additional waveform loss $L_w$.
% The probability distribution of the encoder $q_\phi(w_i|x)$ for all $i$ is trained with joint probability distribution $q_\phi(w|x)$. 
Each loss term is weighted by the corresponding coefficients and then summed up.

We found that a large $\beta_3$ in the early training stage helps the model effectively learn the waveform shape. Hence, we use an exponentially decreasing schedule for its weight $\beta_{\text{w}}$ as follows,
\begin{align}
\beta_{3}(\text{epoch}) &=
\beta_{3}^{(0)} \times (20\exp (-r\times\text{epoch})+1)
\end{align}
where the decrease rate $r$ is $0.144$.

We empirically set the weights of the loss terms to $\beta_1=0.354$, $\beta_\text{2}=2.231$, and $\beta_\text{3}^{(0)}=4.170$. We use a \texttt{LinearLR} learning rate scheduler in \texttt{Pytorch} \cite{paszke2019pytorch} with \texttt{start\_factor} of $1.0$, \texttt{end\_factor} of $0.5$, and \texttt{total\_iters} of $1500$. 
%Some values are rounded for clarity.
We train the model for a total of $5000$ epochs.

\begin{table*}
\centering
\small
\caption{Comparison of generation capacity including reconstruction and descriptor errors in Serum and WaveEdit(WE) datasets. Small models and prior learning are indicated in S and PL. Each metric is the average of elements. 
%Waveforms have been normalized, ensuring the minimum value is -1, and the maximum is 1 for precise comparison. The results indicate that our proposed model excels in waveform generation among explored wavetable generator models.
}

\resizebox{\linewidth}{!}{%
\begin{tblr}{
  width = .8\linewidth,
  colspec = {Q[180] Q[50]Q[50] Q[50]Q[50] Q[50]Q[50] Q[50]Q[50] Q[50]Q[50] Q[50]Q[50] Q[50]Q[50]},
  cells = {c},
  cell{1}{1} = {r=2}{},
  cell{1}{2} = {c=2,r=2}{0.115\linewidth},
  cell{1}{4} = {c=2,r=2}{0.115\linewidth},
  cell{1}{6} = {c=10}{0.576\linewidth},
  cell{2}{6} = {c=2}{0.115\linewidth},
  cell{2}{8} = {c=2}{0.115\linewidth},
  cell{2}{10} = {c=2}{0.115\linewidth},
  cell{2}{12} = {c=2}{0.113\linewidth},
  cell{2}{14} = {c=2}{0.113\linewidth},
  hline{2, 3} = {-}{0.08em},
  hline{1} = {-}{0.15em},
  hline{9} = {-}{0.15em},
}
 Model & Waveform MAE &  & Spectral MSE &  & Descriptor MAE &  &  &  &  &  &  &  &  & \\
 &  &  &  &  & Brightness &  & Richness &  & Fullness &  & Undulation &  & Symmetry & \\
& Serum & WE & Serum & WE & Serum & WE & Serum & WE & Serum & WE & Serum & WE & Serum & WE\\
Baseline & 0.054 & 0.055 & 0.177 & 0.488 & 0.047 & 0.080 & \bf{0.045} & \bf{0.045} & 0.070 & \bf{0.121} & -- & -- & -- & -- \\
WS & \bf{0.005} & \bf{0.035} & \bf{0.018} & \bf{0.221} & \bf{0.026} & \bf{0.068} & 0.103 & 0.189 & \bf{0.019} & 0.161 & \bf{0.015} & \bf{0.046} & \bf{0.183} & \bf{0.730} \\
WS-S & 0.008 & 0.038 & 0.033 & 0.240 & 0.062 & 0.097 & 0.204 & 0.239 & 0.033 & 0.231 & 0.029 & 0.055 & 0.376 & 0.926 \\
WS-PL & 0.017 & 0.038 & 0.117 & 0.250 & 0.050 & 0.074 & 0.126 & 0.203 & 0.079 & 0.185 & 0.026 & 0.049 & 0.454 & 0.811 \\
WS-S-PL & 0.024 & 0.045 & 0.170 & 0.309 & 0.092 & 0.105 & 0.213 & 0.240 & 0.136 & 0.311 & 0.041 & 0.058 & 0.791 & 1.067 \\
\end{tblr}
}
\label{tab:tab1}
\end{table*}
\section{Results}
Our model is evaluated for its capacity and computational efficiency in wavetable generation. If not specified, the evaluation is conducted using the test dataset from Serum and our WS model.
\subsection{Wavetable Generation}
\begin{figure}[t]
 \centerline{
 \includegraphics[width=1\columnwidth]{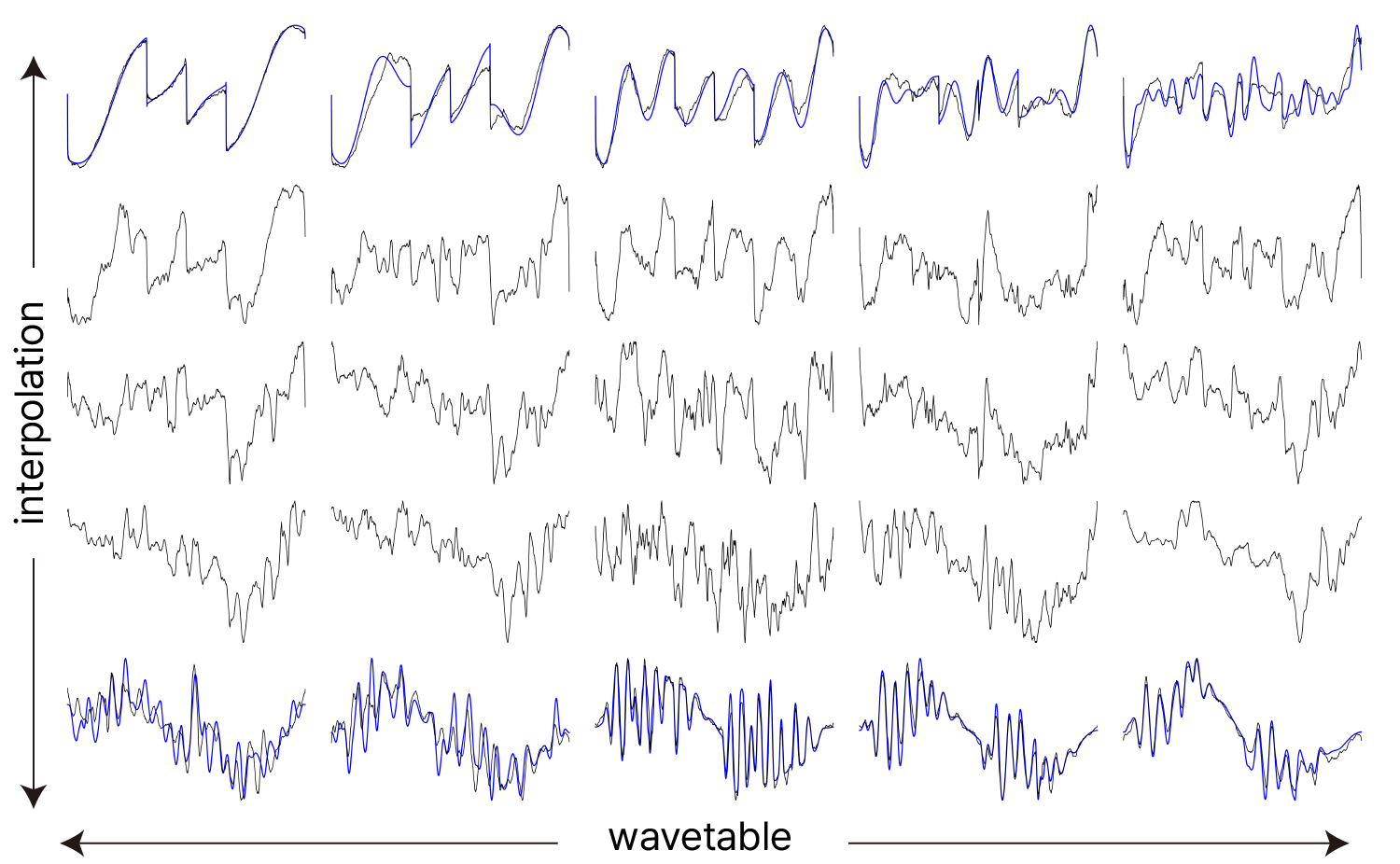}}
 \caption{Latent space interpolation between two styles. Black and blue waveforms are the generated ones and the ground truth, respectively.
 % The latent space interpolation can obtain the continuous waveform transition between individual waveforms in different wavetables.
 %We found that latent values are closely scattered in the corresponding subspace prior. 
 }
 \label{fig:fig3}
\end{figure}

\begin{figure}[t]
 \centerline{
 \includegraphics[width=1\columnwidth]{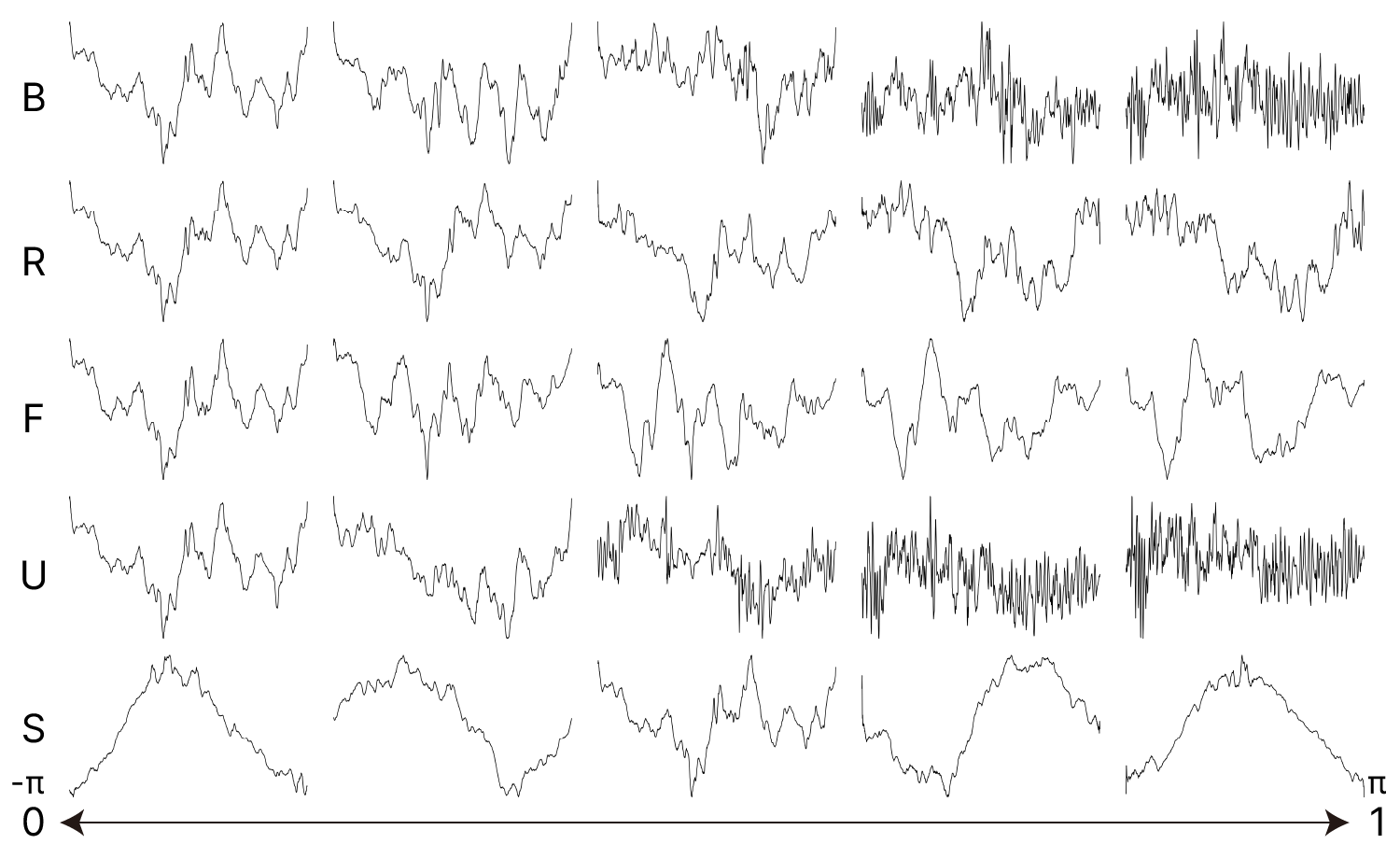}}
 \caption{
 Effect of the descriptors.
 We gradually adjust each descriptor in 5 steps between 0.2 to 1.0 ($-\pi$ to $\pi$ for symmetry), fixing style latents at zero.
 }
 \label{fig:fig4}
\end{figure}

We demonstrate wavetable generation by adjusting parameters in style subspaces and descriptors. We conduct wavetable generation by the latent space interpolation across two styles. We select two different wavetables that represent each style. We excerpt 5 waveforms at equal intervals and pair them sharing the same wavetable position. We encode them with extracted descriptors and linearly interpolate between the paired parameters in the latent space. \autoref{fig:fig3} shows the wavetable generation between two wavetables in different styles. We conclude that the smooth transition between two waveforms of different styles allows us to generate wavetables of unseen waveforms that include novel sonic textures.
We also show how each descriptor parameter morphs the waveform in \autoref{fig:fig4}. The waveforms morph distinctively according to the given descriptor. Notably, our proposed morphological descriptors work along our assumption. 
% The waveform indicates more zigzags as undulation increases and the waveforms represent different phases for the symmetry.
% The points are mapped to the latent space derived from a pair of in-domain wavetables, tracing the trajectory within the associated subspace. The findings revealed that
% each wavetable exhibited a trajectory closely aligned with its conditioned prior.
% Further, we generated a series of 256 wavetables employing interpolation between paired waveforms sharing identical wavetable positions. This observation underscores the continuous interpolation of waveforms between two distinct entities.

\begin{figure}[t]
 \centerline{
 \includegraphics[width=1\columnwidth]{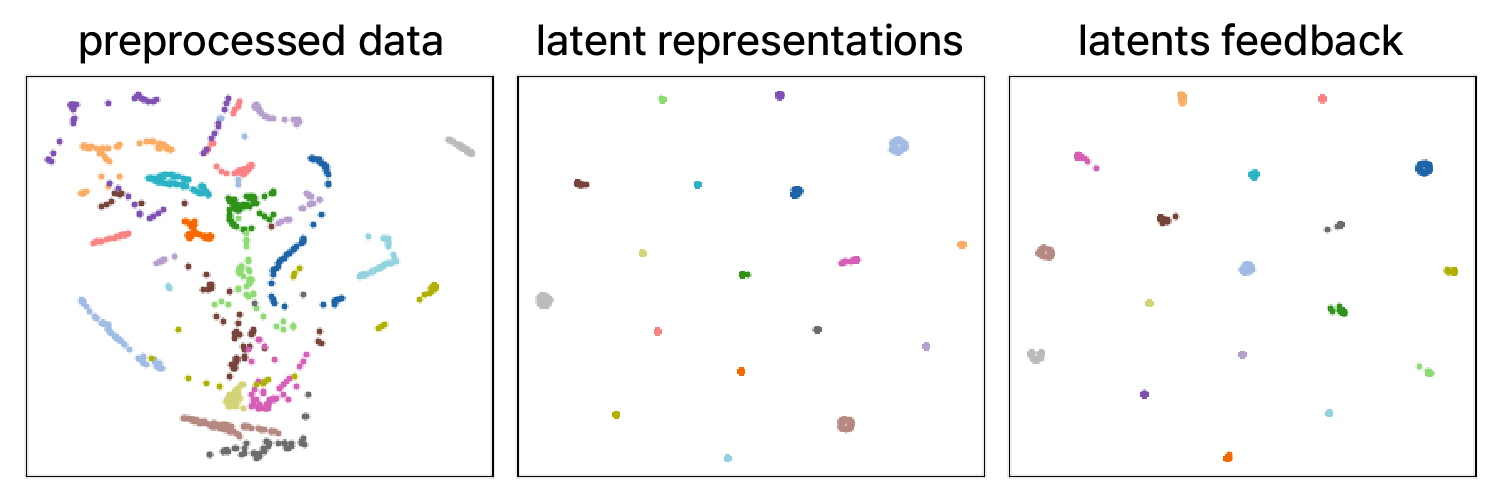}}
 \caption{t-SNE scatter plot of preprocessed test data, the latents, and the feedback latents.
 Different colors represent different wavetables.}
 \label{fig:fig5}
\end{figure}
\subsection{Generation Capacity}
\subsubsection{Latent Space Factorization}
\autoref{fig:fig5} shows the distribution of our data and latent representations. Our model effectively disentangles waveform styles as KL divergence indicated to 0.133, which is lower than 0.393 of the baseline. Hence, the encoder distinguishes the waveform styles better than the previous method. However, our prior learning model shows 25.133 to this metric, indicating the encoder's low performance. As we found that the learning priors are not converging to certain values, we suggest that they hinder learning latent subspaces because the criteria for calculating KL divergence keep changing, leading to unstable training. We also take the analysis of the feedback latents by feeding back reconstructed waveforms into the encoder. The t-SNE plot indicates clear disentanglement of the feedback latents. The KL divergence for feedback latents is 0.116, which is lower than the original latent representation. This demonstrates the decoder's capability to generate waveforms according to a style.
\subsubsection{Reconstruction}
We evaluate our model with mean absolute error (MAE) between the waveform domain and mean square error (MSE) in the magnitude spectrum domain. We calculate descriptor errors by MAE between extracted descriptors from $\hat{x}$ and $x$. The MAE of symmetry $\text{Sym}(\hat{x},x)$ passes an additional operation
\begin{align}
    \text{Sym}^{'}(\hat{x},x) &= -|(\text{Sym}(\hat{x},x) \mod 2\pi) -\pi|+\pi
\end{align}
to avoid a loss bigger than $\pi$ since we only measure the smaller angle.  \autoref{tab:tab1} indicates the reconstruction capacity of the baseline and our models. Our proposed model shows exceptional outperforming in waveform MAE and spectral MSE compared to the baseline. Our models are also competent capacity in the descriptor MAE, especially in brightness. Our models with learnable priors indicate the minimum reconstruction performance. We conclude that training the priors is not the best choice for learning latent subspace. Comparing the results between the two datasets, it is clear that the performance generally excels in the Serum dataset. Thus, reconstruction capacity highly depends on the training dataset, which is analogous to the results in \cite{krekovic2022deep}.
%addaddadada
\begin{table}[t]
\centering
\small
\caption{Ablation study results (Serum dataset). The original model's performance is shown in the top row. Descriptors are indicated in their initials.}

\resizebox{\linewidth}{!}{%
\begin{tblr}{
  width = \linewidth,
  colspec = {Q[22]Q[22]Q[22]Q[54]Q[54]Q[54]Q[54]Q[54]Q[54]Q[54]},
  cells = {c},
  hline{1} = {-}{0.15em},
  hline{2} = {-}{0.08em},
  hline{8} = {-}{0.15em},
  %vline{4} = {-}{0.04em},
  hline{3} = {-}{0.02em},
}

\ding{192}&\ding{193}&\ding{194}&MAE&MSE&\,\,\,\,B&\,\,\,\,R&\,\,\,\,F&\,\,\,\,U&\,\,\,\,S\\
\cmark&\xmark&\xmark& \bf{0.005} & \bf{0.018} & \bf{0.026} & 0.103 & \bf{0.019} & \bf{0.015} & 0.183\\
\xmark&\xmark&\xmark& 0.015 & 0.086 & 0.083 & 0.140 & 0.126 & 0.033 & 0.468\\
\xmark&\xmark&\cmark& 0.022 & 0.147 & 0.116 & 0.171 & 0.178 & 0.047 & 0.686\\
\cmark&\cmark&\xmark& 0.005 & 0.023 & 0.064 & 0.143 & 0.020 & 0.019 & \bf{0.077}\\
\cmark&\xmark&\cmark& 0.012 & 0.037 & 0.040 & 0.112 & 0.054 & 0.022 & 0.490\\
\cmark&\cmark&\cmark& 0.012 & 0.041 & 0.055 & \bf{0.097} & 0.049 & 0.019 & 0.097\\
\end{tblr}
}
\label{tab:tab2}

\ding{192} Descriptors \,\, \ding{193} Descriptor Loss \,\,\ding{194} Spectral Output

\end{table}

\subsection{Ablation Study}
We provide more information about the various settings we have tested, indicating the relationship between our methods and reconstruction capacity. We discuss three elements of our experiment; see \autoref{tab:tab2}.
First, we can get rid of the use of descriptors and the relative losses (\ding{192} in \autoref{tab:tab2}). 
Second, we add $L_1$ descriptor loss with coefficients $\boldsymbol{\beta}_{4} \cdot \boldsymbol{\mathcal{L}}_{\text{d}}$, where $\boldsymbol{\beta}_\text{4}=[4.17, 4.17, 4.17, 10, 40]$. We calculate descriptor loss using the same method as the evaluation metrics (\ding{193} in \autoref{tab:tab2}). 
Lastly, we change the direct waveform output to the spectral components. The spectral components consist of magnitude $|X|$ and phase $\angle X$ where $X$ denotes the DFT of the single waveform $x$. We convert back to the time domain via inverse Fast Fourier Transform (\ding{194} in \autoref{tab:tab3}). We train models with altered settings and evaluate their reconstruction capacity. The results show that our current setup has overall better reconstruction quality than the other alternatives.
% The results show low reconstruction capacity without the concatenation of descriptors.

\subsection{Computational Efficiency}

\begin{table}
\small
\centering
\caption{The decoders' computational efficiency in CPU. Time refers to the generation time per waveform.}
{
    \begin{tblr}{
        colspec ={Q[50]Q[50]Q[60]Q[100]Q[50]},
        cells = {c},
        cell{1}{1} = {r=2}{},
        cell{1}{2} = {r=2}{0.115\linewidth},
        cell{1}{3} = {r=2}{0.115\linewidth},
        cell{1}{4} = {c=2}{},
      hline{1,6} = {-}{0.15em},
      hline{2} = {-}{0.08em},
      hline{3} = {-}{0.08em},
    }
    Model&\# of Params &FLOPs ($\times 10^{6}$)&Computation Speed \\
    &&& Time (ms) & RTF \\ 
    Baseline & 1.2M & 66.902 & 26.2 & 0.81  \\
    WS & 1.2M & 3.443 & 17.2  & 1.24   \\
    WS-S & 76.3K & 0.219 & 9.7  & 2.20   \\
    \end{tblr}
}
\label{tab:tab3}
\end{table}

To use \texttt{Wavespace} in real-time audio plug-ins,
%Since our model works under real-time computation in audio plug-ins, 
it is crucial to check whether the decoder's computational efficiency is within the appropriate generation speed. 
We report floating-point operations (FLOPs) and real-time factor (RTF) for averaged 100 times of dataset in \autoref{tab:tab3}.
We calculate the RTF as processing time divided by buffer duration. 
Assuming that we choose 1024 buffer length with a 48000Hz sample rate, the buffer duration is approximately 21.3 milliseconds.
We utilize the \texttt{M1} CPU chip of the 2020 version \texttt{MacBook}. Therefore, our proposed model's decoder performs real-time generation under typical buffer duration settings.
\section{Plug-in implementation}
\begin{figure}[t]
 \centerline{
 \includegraphics[width=1\columnwidth]{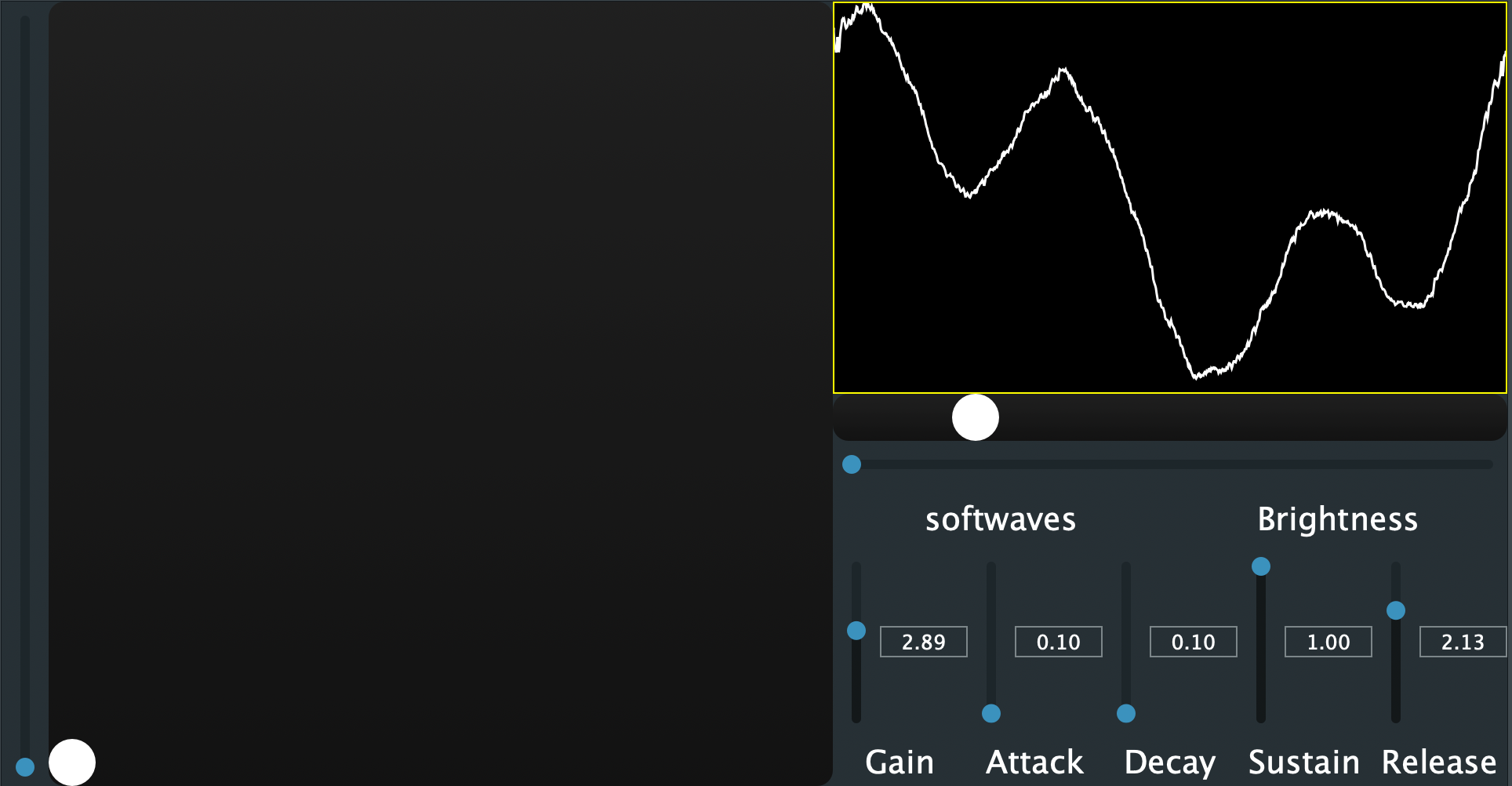}}
 \caption{The snapshot of our VST plug-in implemented based on \texttt{Wavespace}.}
 \label{fig:fig6}
\end{figure}

{
\iffalse
For the proof of concept for the real-time plug-in in a Steinberg Virtual Studio Technology (VST) format, we prototype a playable VST plug-in leveraging Wavespace. In implementing a real-time virtual instrument with the pre-trained model, finding an appropriate policy for requesting every new waveform is crucial. The model may generate the waveform either when the host requests the next audio buffer or when one of the parameters changes.
% However, each approach struggles with a high sample rate and frequent value changes i.e. parameter automation. Therefore, in our actual implementation, we address this issue by executing the model when condition values change. However, we restrict it if a certain number of samples has not been played since the last waveform generation.
\autoref{fig:fig6} shows the user interface of the virtual instrument. the rectangle on the left shows a style subspace that switches according to the attached slider. On the right is a visualization of the waveform, a small slider for selecting timbral and morphological descriptors, and a big slider for adjusting its value by tweaking back and forth. Users can select a point in each subspace. Source code for our VST plugin is also available online\footnote{\url{https://github.com/kimgihong2510/WSSInst/}.}.
\input{figures/plugin}
\fi
} % 기존 글
% Since the generation time of a waveform is shorter than the typical buffer duration, Simply put, a waveform is generated in every buffer.
We have proved that our model's RTF quantitatively grants real-time capacity.
% To demonstrate this, we implemented a synthesizer that utilizes this method.
As a proof of concept for real-time use, we prototype an autoencoder oscillator that leverages our \texttt{Wavespace}. 
When implementing a real-time virtual instrument with the pre-trained model, finding an appropriate policy for requesting every new waveform is crucial. 
To make it stable under buffer constraints, we (i) execute the model only when parameters change and (ii) restrict the maximum frequency of the model executions.
We represent the user interface of our plugin in \autoref{fig:fig6}.
The left rectangle shows a style subspace $W_i$, and we can switch it to another one using the attached slider. 
Users can select a point $w_i$ in each subspace $W_i$. 
On the right is a visualization of the waveform $x_i$; it has a small slider for selecting a timbral or morphological descriptor and a large slider for adjusting its value. 
Finally, as a basic functionality, we add a gain fader and amplitude envelope with attack, decay, sustain, and release sliders (bottom right).
Our plugin is implemented in virtual studio technology (VST) and audio unit (AU) format and will be available online.
%\footnote{\url{https://github.com/kimgihong2510/WavespaceImplementation}.}. 

\section{Conclusion}
Our study introduced \texttt{Wavespace}, a wavetable generator framework with factorized latent space, allowing users to generate wavetables from scratch. By learning latent subspaces in the CVAE structure, our model efficiently generated waveforms corresponding to the given styles and descriptors. 
We achieved improved reconstruction quality in both waveform and spectral domains. 
Our model is also computationally efficient, enabling the implementation of a real-time autoencoder-based oscillator.

A crucial question for future work is to identify which factors within the dataset significantly contribute to the model's generation capacity. Also, future research should investigate methods for combining parameters in a user-friendly manner, allowing users to leverage their control over wavetable generation efficiently. We hope for further discussions on wavetable generation methods and their creative implementation to audio plug-ins.
%\textcolor{blue}{Limitation and future work include ...}

%실제 사용성이 괜찮은지
\section{Ethics statement}
 We used wavetables in the Serum, a commercial wavetable synthesizer provided by Xfer Records. The company informed us that there are no legal guidances for the license using Serum wavetables beyond music making and sound design. Thus, we did our best to prevent any damage our research may cause to Xfer Records in terms of usage.
 Our research never shares with anyone (i) any implementation of Serum and (ii) raw wavetable datasets from Serum.
%\section{Acknowledgments}
%Do we have any acknowledgment? This work was supported by ... -> 다 저자로 올라가서 없음

%\newpage
\bibliography{ref}
\end{document}